\documentclass{desyproc}

\begin{document}
\title{Status of the ADMX and ADMX-HF experiments}

\author{{\slshape Karl van Bibber$^1$ and Gianpaolo Carosi$^{2}$}\\[1ex]
$^1$University of California, Berkeley, CA, USA\\
$^2$Lawrence Livermore National Laboratory, Livermore, CA, USA}

\contribID{familyname\_firstname}

\desyproc{DESY-PROC-2012-04}
\acronym{Patras 2012} 
\doi  

\maketitle

\begin{abstract}
The Axion Dark Matter eXperiment (ADMX) is in the midst of an upgrade to reduce its system noise temperature. ADMX-HF (High Frequency) is a second platform specifically designed for higher mass axions and will serve as an innovation test-bed. Both will be commissioning in 2013 and taking data shortly thereafter. The principle of the experiment, current experimental limits and the status of the ADMX/ADMX-HF program will be described. R\&D on hybrid superconducting cavities will be discussed as one example of an innovation to greatly enhance sensitivity.
\end{abstract}

\section{Introduction}

A light axion represents a compelling dark-matter candidate, and as demonstrated by Sikivie thirty years ago, such axions could be detected by their conversion to photons in a microwave cavity permeated by magnetic field \cite{Sikivie:1983}, with the signal power given by:

\begin{eqnarray}
P_{Sig} = g^2_{a\gamma\gamma}\left(\frac{\rho_a}{m_a}\right) B^2_0 V C Q.
\end{eqnarray}

\begin{figure}[hb]
\centerline{\includegraphics[width=0.95\textwidth]{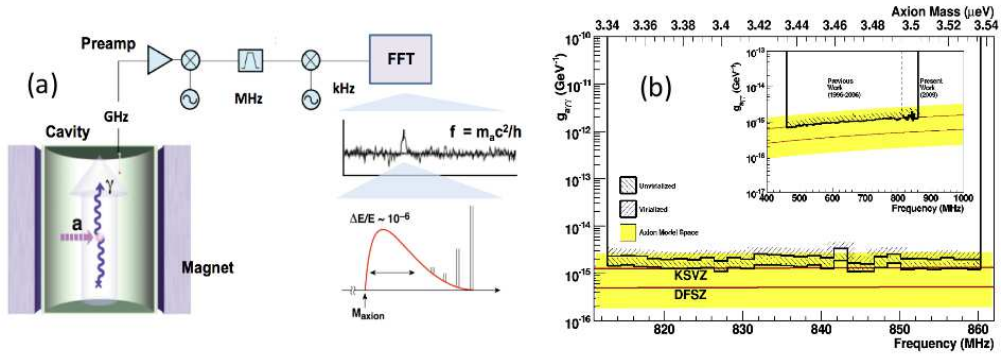}}
\caption{(a) Schematic of the microwave cavity technique. (b) Exclusion region to date by ADMX \cite{Asztalos:2010}}\label{Figure: 1}
\label{fig:1}
\end{figure}

The signal is expected to be exceedingly weak, $< 10^{-21}$ W.  Pioneering experiments at Brookhaven \cite{DePanfilis:1987} and Florida \cite{Hagmann:1990} established the basic design concept (Figure \ref{fig:1}a) based on a superconducting solenoidal magnet, a tunable high-Q copper cavity, and state-of-the art cryogenic amplifiers as the front end of a heterodyne receiver.  The experiment is step-tuned, and after an integration period of $\sim$ 100 sec., the FFT spectrum within the cavity bandpass is calculated.  The Axion Dark Matter eXperiment (ADMX) represented a large scale up from the first generation experiments (the 8T solenoid is 1.1m long and 60cm in diameter), and marked the beginning of a long campaign to reduce the system noise temperature. ADMX has published an exclusion region over roughly an octave in mass (460-860 MHz, or 1.9-3.6 $\mu$eV) for KSVZ axions saturating the galactic halo (Figure \ref{fig:1}b) \cite{Asztalos:2010}. ADMX has developed and deployed Microstrip-coupled SQUID Amplifiers (MSA) \cite{Muck:1998} which have demonstrated near quantum-limited performance in the laboratory. 

\section{ADMX upgrade}

The ADMX experiment ran with a MSA from 2008-2010 at pumped LHe temperatures ($\sim$1.2 K). The experiment was then moved in the summer of 2010 from LLNL to the University of Washington (UW) where it is currently being rebuilt to include a dilution refrigerator. This will allow for operations at a physical temperature of $\sim$100 mK, a regime in which the MSA is expected to be quantum limited. An initial data-run with a pumped $^3$He refrigerator ($T_{phys} \approx 400\;mK$) will begin at the end of the summer of 2013 and run until it can be replaced with the colder dilution refrigerator (spring 2014). In addition to the lowered system noise temperature a second antenna and receiver system is being added in order to take data with higher order TM modes (such as the TM$_{020}$). These modes move in parallel to the fundemental TM$_{010}$ mode and have lower, but non-negligible, coupling to axions. This will allow the ADMX experiment to search for axions in two frequencies at the same time, greatly increasing the detection potential. Additional improvements include revamped microwave cavities, motion control systems and an updated receiver chain that takes advantage of new digital electronics. The primary participants in the ADMX Generation 2 experiment are UW, LLNL, UC Berkeley, Univ. Florida, NRAO, and Sheffield University. 

\section{ADMX-HF}

To focus on specific challenges of the axion search at high masses, and thus high frequency (the resonant condition being $h\nu = m_{a}c^2$), and significantly improving the scan rate, we are building a second smaller platform, called ADMX-HF (High Frequency) designed for the 4 - 40 GHz range.  This experiment is being built up by a collaboration of Yale (the host),  JILA/Colorado, UC Berkeley and LLNL.  Supported by NSF, this platform will both produce data of intermediate sensitivity (KSVZ model), and serve as an innovation platform, to allow rapid testing of new receiver and cavity concepts, etc. 

Tailored to higher frequencies, ADMX-HF is physically smaller than its lower frequency counterpart.  The superconducting magnet is a solenoid of only 15 cm x 40 cm; it has a 9T central field and was designed to have a radial component limited to $B_{r} < 50 G$, permitting operation with microwave cavities with Type II superconducting thin films as the barrel section (see next section).  The entire experiment is cooled by a dilution refrigerator to 25 mK.  

\begin{figure}[hb]
\centerline{\includegraphics[width=0.9\textwidth]{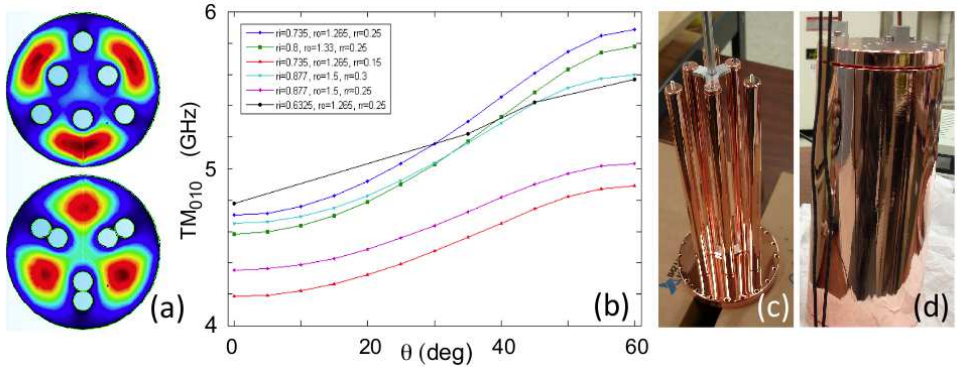}}
\caption{(a) Stator-rotor tuning scheme of the microwave cavity.  The false-color indicates the magnitude of the axial electric field of the TM$_{010}$ mode.  (b)  The frequency of the TM$_{010}$ mode as a function of tuning angle.  Photographs of the cavity (c) with, and (d) without the cavity barrel and top end-cap removed.}\label{Figure: 2}
\label{fig:2}
\end{figure}

Figures \ref{fig:2}a,b demonstrate the tuning of the TM$_{010}$ mode between 4.6 - 5.9 GHz by rotation of three axial rods with respect to three fixed rods. The microwave cavity is OFHC copper electrodeposited on stainless steel and annealed (Figures \ref{fig:2}c,d); at 4K a quality factor Q $\sim$ 40,000 is expected, for the complete cavity, critically coupled.

\begin{figure}[hb]
\centerline{\includegraphics[width=0.9\textwidth]{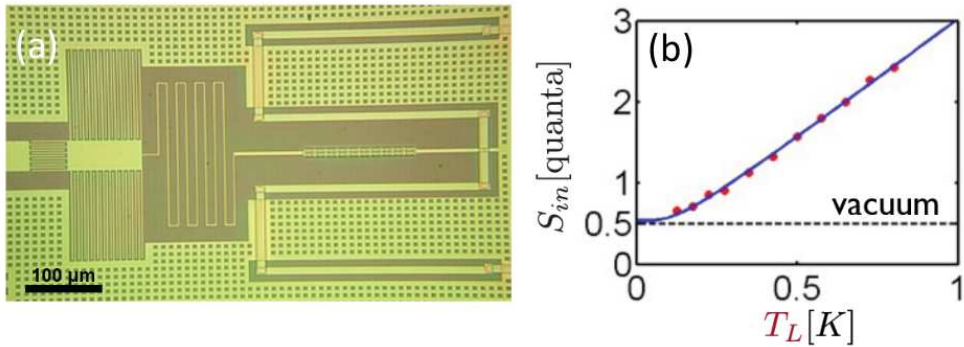}}
\caption{(a) Microphotograph of the Josephson Parametric Amplifier.  (b)  Demonstration of quantum-limited performance at temperatures below 100 mK. }\label{Figure: 3}
\label{fig:3}
\end{figure}

One of the innovations that has been incorporated is the use of Josephson Parametric Amplifiers (JPA).  The initial run will use a JPA developed by K. Lehnert of JILA [6,7] and which has already seen use in studies of quantum nanomechanical oscillators, and quantum information (Figure \ref{fig:3}a).  This device is continuously tunable from 4-8 GHz, achieving quantum-limited performance (Figure \ref{fig:3}b), even after the HEMT post-amplifier, owing to its very high gain ($>$30 dB).  The JPA can also be operated in a squeezed-state mode to achieve sub-quantum limited noise, not an objective for the first data run.    

\section{Hybrid superconducting cavity R\&D}

One of the current R\&D efforts underway in the ADMX collaboration is to improve the Q factor of the microwave cavities used. The linewidth of gravitationally virialized axions is expected to be $\sim\beta^2 \sim(10^{-3}c)^2 \sim10^{-6}$, which is equivelant to a $Q_a = E/\Delta{E} \sim 10^6$, which is approximately an order of magnitude larger than the $Q_{cavity} \sim 10^5$ of standard copper plated cavities. Superconducting cavities, however, can achieve $Q>10^{10}$ but are driven normal in the presence of a strong magnetic field. One possible solution is to replace the parallel walls of the cavity with a thin film superconductor that can maintain its superconducting properties in the presence of a strong parallel magnetic field. Evidence for this has been seen in NbTiN films up to 10 Tesla \cite{Xiaoxiang}. The Q would then be dominated by the contribution from the regular copper endcaps to the cavity which should give an enhancement of $Q_{hybrid} = (1 + L/R)Q_{Cu}$, where $L$ and $R$ is the length and radius of the cavity and $Q_{Cu}$ is for an all-copper cavity. Currently these R\&D efforts are taking place at LLNL, Yale and Univ. Florida.

\section{Summary and conclusions}

The microwave cavity dark matter axion experiment now has a clear path both in achieving the requisite sensitivity (DFSZ models and below), and mass reach, initially up to 100 $\mu$eV.  Extending the search to much higher masses will require the introduction of hybrid superconducting cavities, and different receiver technologies, possibly bolometers.  However, the ADMX and AMDX-HF platforms will already be in prime discovery territory this year.

\section{Acknowledgments}

This work was performed under the auspices of the U.S. Department of Energy by Lawrence Livermore National Security, LLC, Lawrence Livermore National Laboratory under Contract DE-AC52-07NA27344 and under NSF grants PHY-1067242, and PHY-1306729. LLNL-PROC-635756.

\section{Bibliography}
 

\begin{footnotesize}

\end{footnotesize}


\end{document}